\documentclass[twocolumn,pra,superscriptaddress]{revtex4}
\usepackage{amsfonts}
\usepackage{amssymb}
\usepackage{amsmath}
\usepackage{epsfig}
\usepackage{color}
\usepackage{graphics, graphicx}
\usepackage{bbold}
\usepackage{psfrag}
\usepackage{mathcomp}
\usepackage{subfigure}
\usepackage{verbatim}
\usepackage{float}
\usepackage{graphicx}
\usepackage[colorlinks,citecolor=blue]{hyperref}
\makeatletter

\newcommand{\Rmnum}[1]{\expandafter\@slowromancap\romannumeral #1@}
\makeatother

\begin{document}

\title{Generalized Aubry-Andr\'{e}-Harper Models in Optical Superlattices}

\author{Yi Li}
\affiliation{State Key Laboratory of Quantum Optics and Quantum Optics Devices, Institute
of Laser Spectroscopy, Shanxi University, Taiyuan, Shanxi 030006, China}
\affiliation{Collaborative Innovation Center of Extreme Optics, Shanxi
University,Taiyuan, Shanxi 030006, China}

\author{Jia-Hui Zhang}
\affiliation{State Key Laboratory of Quantum Optics and Quantum Optics Devices, Institute
of Laser Spectroscopy, Shanxi University, Taiyuan, Shanxi 030006, China}
\affiliation{Collaborative Innovation Center of Extreme Optics, Shanxi
University,Taiyuan, Shanxi 030006, China}

\author{Feng Mei}
\email{meifeng@sxu.edu.cn}
\affiliation{State Key Laboratory of Quantum Optics and Quantum Optics Devices, Institute
of Laser Spectroscopy, Shanxi University, Taiyuan, Shanxi 030006, China}
\affiliation{Collaborative Innovation Center of Extreme Optics, Shanxi
University,Taiyuan, Shanxi 030006, China}

\author{Jie Ma}
\affiliation{State Key Laboratory of Quantum Optics and Quantum Optics Devices, Institute
of Laser Spectroscopy, Shanxi University, Taiyuan, Shanxi 030006, China}
\affiliation{Collaborative Innovation Center of Extreme Optics, Shanxi
University,Taiyuan, Shanxi 030006, China}

\author{Liantuan Xiao}
\affiliation{State Key Laboratory of Quantum Optics and Quantum Optics Devices, Institute
of Laser Spectroscopy, Shanxi University, Taiyuan, Shanxi 030006, China}
\affiliation{Collaborative Innovation Center of Extreme Optics, Shanxi
University,Taiyuan, Shanxi 030006, China}

\author{Suotang Jia}
\affiliation{State Key Laboratory of Quantum Optics and Quantum Optics Devices, Institute
of Laser Spectroscopy, Shanxi University, Taiyuan, Shanxi 030006, China}
\affiliation{Collaborative Innovation Center of Extreme Optics, Shanxi
University,Taiyuan, Shanxi 030006, China}

\date{\today}
\begin{abstract}
  Ultracold atoms trapped in optical superlattices provide a simple platform for realizing the seminal Aubry-Andr\'{e}-Harper (AAH) model. However, the periodic modulations on the nearest-neighbour hoppings have been ignored in this model. In this paper, we find that optical superlattice system actually can be approximately described by a generalized AAH model in the case of $V_1\gg V_2$, with periodic modulations on both on-site energies and nearest-neighbour hoppings, supporting much richer topological properties that are absent in the standard AAH model. Specifically, by calculating Chern numbers and topological edge states, we show that the generalized AAH model possesses multifarious topological phases and topological phase transitions, as compared to the standard AAH model only supporting a single topological phase. Our findings can open up more opportunities for using optical superlattices to study topological and localization physics.
\end{abstract}

\maketitle

Ultracold atoms in optical lattices in the past decade have been proved to be a powerful platform for exploring topological phases of matter~\cite{TPCArev1,TPCArev2,TPCArev3}. This platform features unprecedented controllability and flexibility, opening possibilities to go beyond standard solid-state topological systems. For example, ultracold atoms allow to create a synthetic dimension inside~\cite{Lewenstein2012,Lewenstein2014}. The basic idea is that by coupling a sets of atomic states in a sequential manner, ranging from internal hyperfine~\cite{Fallani2015,Spielman2015}, magnetic~\cite{Nascimbene2020}, clock~\cite{Ye2016,Fallani2016,Ye2017} and Rydberg states~\cite{Killian2022}, to external momentum~\cite{Gadway2016,Ma2022}, orbital~\cite{Shin2020} and superradiant states ~\cite{sms,Chen2018,Wang2019,Wang2021,Zhang2021}, one can construct a synthetic dimension in which the atomic states are treated as lattice sites and the couplings between them as the lattice hoppings. This approach enables the implementation of high-dimensional topological models in low-dimensional optical lattices with a synthetic real-space dimension, and also provides opportunities for exploring the topological effects that remain challenging before, such as the realization of chiral edge states in optical lattices~\cite{Fallani2015,Spielman2015} and Laughlin's topological pump~\cite{Nascimbene2021}.

The concept of synthetic dimension also can be generalized to momentum space. Taking the advantage of high controllability in optical lattices, one can regard a periodic systematic parameter as an extra momentum, and by tuning such parameter from $0$ to $2\pi$ a synthetic momentum space is built. Recent studies have shown that, optical superlattices offer a natural platform for realizing such synthetic dimension~\cite{Chen2012,Mei2012}. This sort of superlattice is created by superimposing a long optical lattice on a short one, both produced by standing-wave lasers. The corresponding tight-binding Hamiltonian is naturally described by the seminal AAH model~\cite{Harper,AA}. As the two standing-wave lasers have incommensurate wavelengths, the incommensurate AAH model~\cite{Harper,AA} that is well known for displaying Anderson localization transition~\cite{AA,Grempel1982,Modugno2009,
Biddle2009,Biddle2011,Dalfovo2011,Zhang2013,Chen2014,Lxp2015,Lxp2016,Gao2016,Chen2017,Li2017,Li2020,Liu2020} can be implemented, enabling observation of Anderson localization of Bose-Einstein condensate~\cite{Inguscio2008}. In the commensurate case, this one-dimensional model supports multiple energy bands and can be mapped to two dimensions by associating the relative phase of the two standing-wave lasers as a synthetic momentum, allowing us to study the topological properties of two-dimensional integer quantum Hall insulator~\cite{Chen2012,Kraus2012} and $Z_2$ topological insulator phases~\cite{Mei2012}.

We notice that, previous works have been focusing on an extreme case where the periodic modulations on the nearest-neighbour hoppings have been ignored, which leads to wanted standard AAH model with only on-site periodic modulations~\cite{Chen2012,Mei2012,Inguscio2008}. In this paper, we highlight that in the limiting case $V_1\gg V_2$, the corresponding optical superlattice system realizes a generalized AAH model, with periodic modulations on both on-site energies and nearest-neighbour hoppings. We present the detailed derivation for this model, including from the single-particle Hamiltonian to the approximated tight-binding Hamiltonian. By seeing the relative laser phase as a synthetic momentum and considering the commensurate case, the one-dimensional generalized AAH model can be mapped to a two-dimensional lattice model describing a generalized integer quantum Hall effect. Based on calculating Chern numbers and topological edge states, we demonstrate that this model holds much richer topological properties that are absent in the standard AAH model, including multiple topological phases and different forms of topological edge states.

Compared to previous works on off-diagonal AAH models~\cite{AAH1,AAH2,AAH3}, our study shows that the generalized AAH model presented here could support multiple multiband topological phases and various topological phase transitions, which has not been reported before. Particularly for the case when the AAH models support even numbers of energy bands, our work shows that the middle two bands are corresponding to nontrivial gapped topological phases, while the two middle bands in previous works are gapless topological phases~\cite{AAH1,AAH2,AAH3}. Moreover, extending the generalized AAH model to the incommensurate case could offer opportunities beyond the standard AAH model for studying its underlying localization features~\cite{AAHL1,AAHL2}.

\begin{figure}[htb]
	\centering
	\includegraphics[width=8.8cm,height=2.3cm]{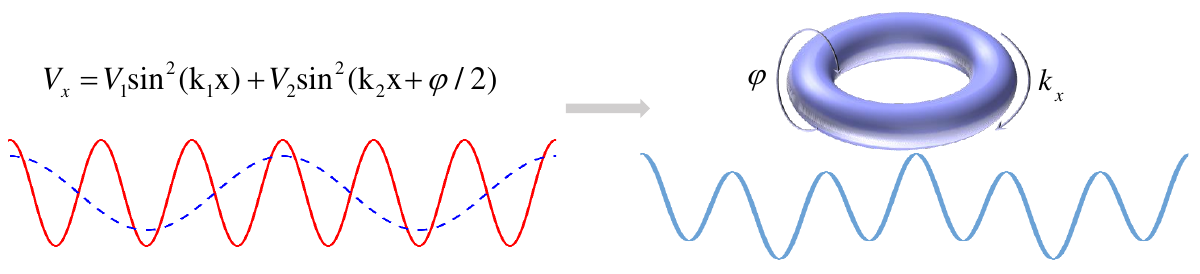}
	\caption{Schematic illustration of the optical superlattice created by superimposing two standing-wave lasers. By regarding the relative laser phase $\varphi$ as a synthetic dimension, such one dimensional lattice allows to explore nontrivial topological properties attributed to systems in two dimensions. }
\label{Fig1}
\end{figure}

\emph{Generalized Aubry-Andr\'{e}-Harper models in optical superlattices}. The optical superlattice with ultracold atom trapped inside is simply created by superimposing a short and a long optical lattice, as shown in Fig. \ref{Fig1}, where the two optical lattice potentials are respectively produced by two standing-wave lasers. The resulted optical superlattice potential takes the following form
\begin{equation}
	V_x=V_1\sin^2(k_1x)+V_2\sin^2(k_2x+\varphi/2)
\end{equation}
where $V_{1,2}$ are respectively the strengths of the short and long optical lattice potentials, $k_{1,2}=2\pi/\lambda_{1,2}$ are the corresponding laser wave vectors and $\varphi$ is the relative phase of the two lasers. Throughout this work, we assume $V_1$ is larger than $V_2$, so that the short lattice is the primary lattice, determining the period of the whole optical superlattice $a=\lambda_1/2$, and the long lattice is a perturbation, causing modulations to the nearest-neighbour hoppings and on-site energies.

The single-particle Hamiltonian for the short optical lattice system is written as
\begin{equation}
\hat H_{s1}=\frac{p_x^2}{2m}+V_1\sin^2(k_1x).
\label{H_s}
\end{equation}
In the second quantization, the continuum single-particle Hamiltonian $H_s$ is written into
\begin{equation}
 \hat H_1= \int {dx{\psi ^ + }\left( x \right){\hat H_{s1}}\psi \left( x \right)}
 \label{Hf}
\end{equation}
Here we assume atoms trapped in the ground band. The field operator is expanded as
\begin{equation}
	\psi \left( x \right) = \sum\limits_m {{c_m}W\left( {x - {x_m}} \right)},
\label{wf}
\end{equation}
where $c_m$ is the annihilation operator at the lattice site $x_m$ and $W\left({x-{x_m}}\right)$ is the corresponding ground-band Wannier function. The lattice spacing is assumed as $a=1$ throughout this work. By substituting Eq. (\ref{wf}) into Eq. (\ref{Hf}), we obtain the tight-binding Hamiltonian
\begin{equation}
	\hat H_1 = \sum\limits_m t_{0}\left( c_m^\dag c_{m +1} + \text{H.c.} \right),
\label{H_1}
\end{equation}
where
\begin{equation}
t_0=\int dx W^*(x-x_{m})(\frac{p^2_x}{2m}+V_1\sin^2{(k_1x)})W(x-x_{m+1}).
\label{t0}
\end{equation}

The single-particle Hamiltonian for the long lattice is
\begin{equation}
\hat H_{s2}=V_2\sin^2(k_2x+\varphi/2).
\label{H_s}
\end{equation}
As $V_1\gg V_2$, the long lattice does not substantially change the minimal position of the short lattice. Then, the field operator still can be approximately expanded in the basis of the Wanneir states defined in terms of the short lattice. With this approximation, we can easily obtain the expression of the tight-binding Hamiltonian for the whole optical superlattice system. Via the same procedure for obtaining $H_1$, the tight-binding Hamiltonian for $H_{s2}$ can be approximately expressed as
\begin{equation}
	\hat H_2 = \sum\limits_m [{t_{m,m+1}\left( {c_m^\dag {c_{m +1}} + \text{H.c.}} \right) + \Delta_m c_m^\dag {c_m}}].
\label{H_2}
\end{equation}
As shown, the introduction of the weak long lattice can cause modulations to both the nearest-neighbour hoppings and on-site energies. Specifically, the modulation to the nearest-neighbor hopping rates are approximately given by
\begin{align}
t_{m,m+1}=&\int dx W^*(x-x_{m})H_{s2}W(x-x_{m+1})  \nonumber \\
=-&\frac{V_2}{2}\int dxW^*(x-x_{m})\cos(2k_2x+\varphi)W(x-x_{m+1}) \nonumber\\
=&-t_1\cos \left( 2\pi\beta m+\varphi \right) + t_2\sin \left( 2\pi\beta m +\varphi \right),
\label{t}
\end{align}
with $\beta=k_2/k_1$ being the commensurability parameter,
\begin{align}
&t_1 = \frac{V_2}{2}\int dxW^*\left( x\right)\cos \left( 2\beta k_1x \right) W\left( x- 1 \right),   \nonumber \\
&t_2 = \frac{V_2}{2}\int dxW^*\left( x \right)\sin \left( 2\beta k_1x \right) W\left( x-1 \right).
\label{t12}
\end{align}
The modulation to the on-site energies are approximately given by
\begin{align}
\Delta_{m}=&\int dxW^*(x-x_m)H_{s2}W(x-x_m)  \nonumber \\
=V_2\int dx W^*&(x-x_m) \dfrac{1-\cos(2k_2x+\varphi)}{2}W(x-x_m)  \nonumber \\
=&\Delta\cos(2\pi\beta m+\varphi)+\text{c.e.},
\label{delta}
\end{align}
with
\begin{align}
\Delta=-\frac{V_2}{2}\int dx W^*(x)\cos(2\beta k_1 x)W(x),
\label{t12}
\end{align}
where $\text{c.e.}$ denotes a constant energy which can be safely neglected without affect main physics.

\begin{figure*}[htb]
	\centering
	\includegraphics[width=13cm,height=8cm]{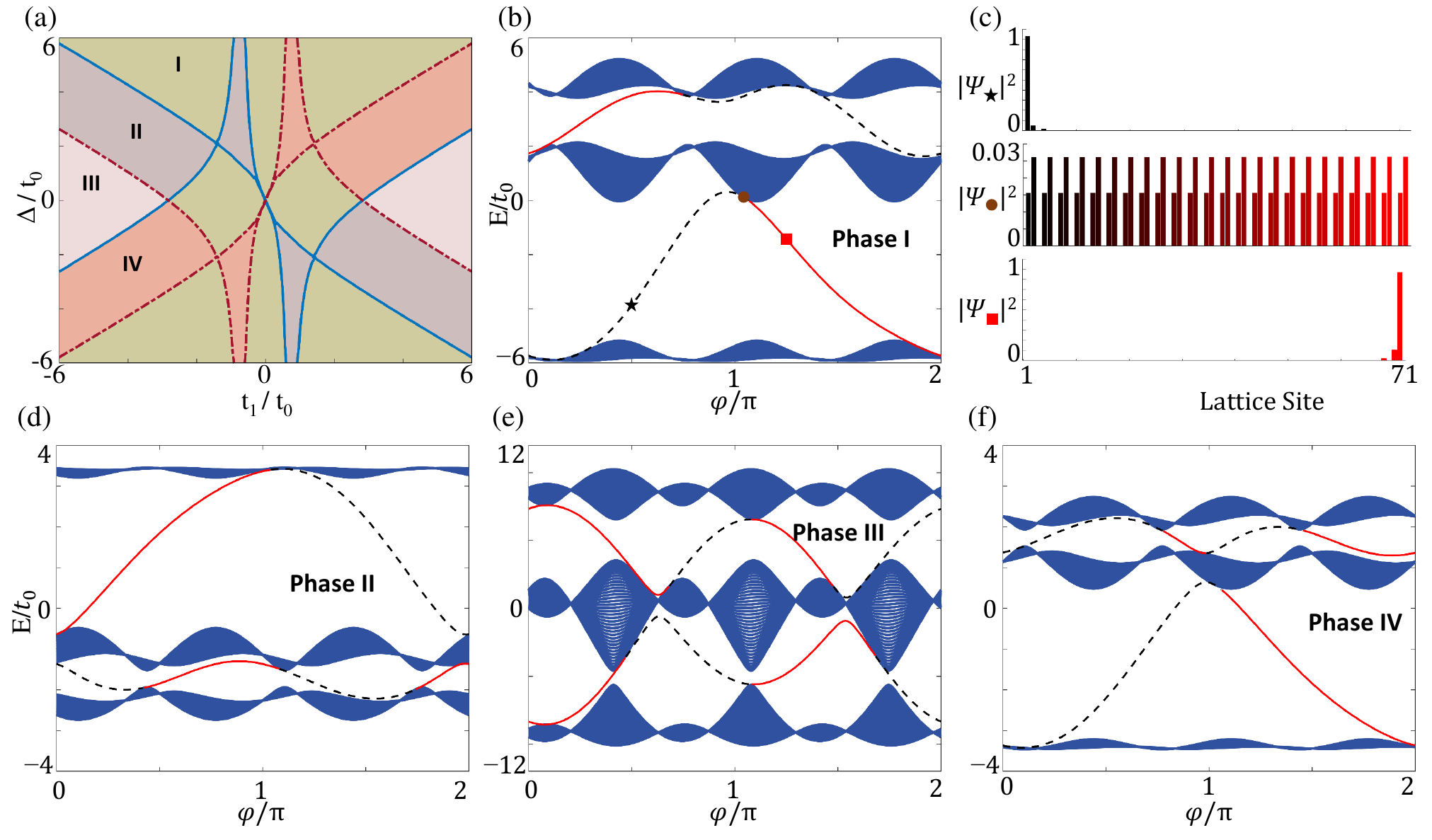}
	\caption{(a) Topological phase diagram in the parameter space of $t_1$ and $\Delta$ for $\beta=1/3$. The
topological phase transitions occur at the closing of the first (blue solid line) or second (red dash-dotted line) energy gaps. The energy spectra of edge states for the generalized AAH model in the phase I-IV are shown in (b) and (d-f), respectively. (c) The density distributions for the three modes marked in (b), showing that the black dashed and red solid lines in (b, d-f) are respectively corresponding to the left and right edge states. The parameters used in the numerical calculations are (b) $t_1=2t_0,\Delta=4t_0$, (d) $t_1=-t_0,\Delta=2t_0$, (e) $t_1=5t_0,\Delta=t_0$ and (f) $t_1=t_0,\Delta=2t_0$. The other parameter is $t_1=t_2$ and $t_0$ is used as the energy unit. }
\label{Fig2}
\end{figure*}

Combining Eqs. (\ref{H_1}) and (\ref{H_2}), we find that the optical superlattice system in the tight-binding limit naturally implements the generalized AAH model,
\begin{equation}
	\hat H_{\text{GAAH}} = \sum\limits_m [(t_0+V^{od}_m)\left( c_m^\dag c_{m+1} + \text{H.c.} \right)+V^d_m c^\dag_m c_m].
\end{equation}
which contains both off-diagonal and diagonal modulations, i.e., $V^{od}_m=-t_1\cos \left( 2\pi\beta m+\varphi \right) + t_2\sin \left( 2\pi\beta m +\varphi \right)$ and $V^{d}_m=\Delta\cos(2\pi\beta m+\varphi)$. It is wroth pointing out that, the derivation of $H_{\text{GAAH}}$ is based on assuming $V_1\gg V_2$ and using approximated Wannier functions, then $H_{\text{GAAH}}$ is not the exact Hamiltonian. However, it can be used to capture the main physical features.

Previous studies have been focusing on an extreme case where the periodic modulations on the nearest-neighbour hoppings have been ignored~\cite{Chen2012,Mei2012,Inguscio2008}, which leads to the standard AAH model Hamiltonian
\begin{equation}
	\hat H_{\text{AAH}} = \sum\limits_m [t_{0}\left( c_m^\dag c_{m+1} + \text{H.c.} \right)+V^d_m c_m^\dag c_m].
\end{equation}
which contains only diagonal modulations. By associating the laser phase $\varphi$ with the momentum $k_y$, the commensurability paramter $\beta$ with the magnetic flux, the standard AAH model can be exactly mapped to the two-dimensional lattice model describing the integer quantum Hall (IQH) effects, i.e,
\begin{equation}
	\hat H_{\text{IQH}} = \sum\limits_{m,n} [t_{0}c_{m,n}^\dag c_{m+1,n}
      +t_ye^{i2\pi\beta m}c_{m,n}^\dag c_{m,n+1}+\text{H.c.}],
      \label{IQH}
\end{equation}
where $t_y=\Delta/2$. In this way, the standard AAH model inherits the topological properties of integer quantum Hall states.

\begin{figure*}[htb]
	\centering
	\includegraphics[width=17cm,height=8cm]{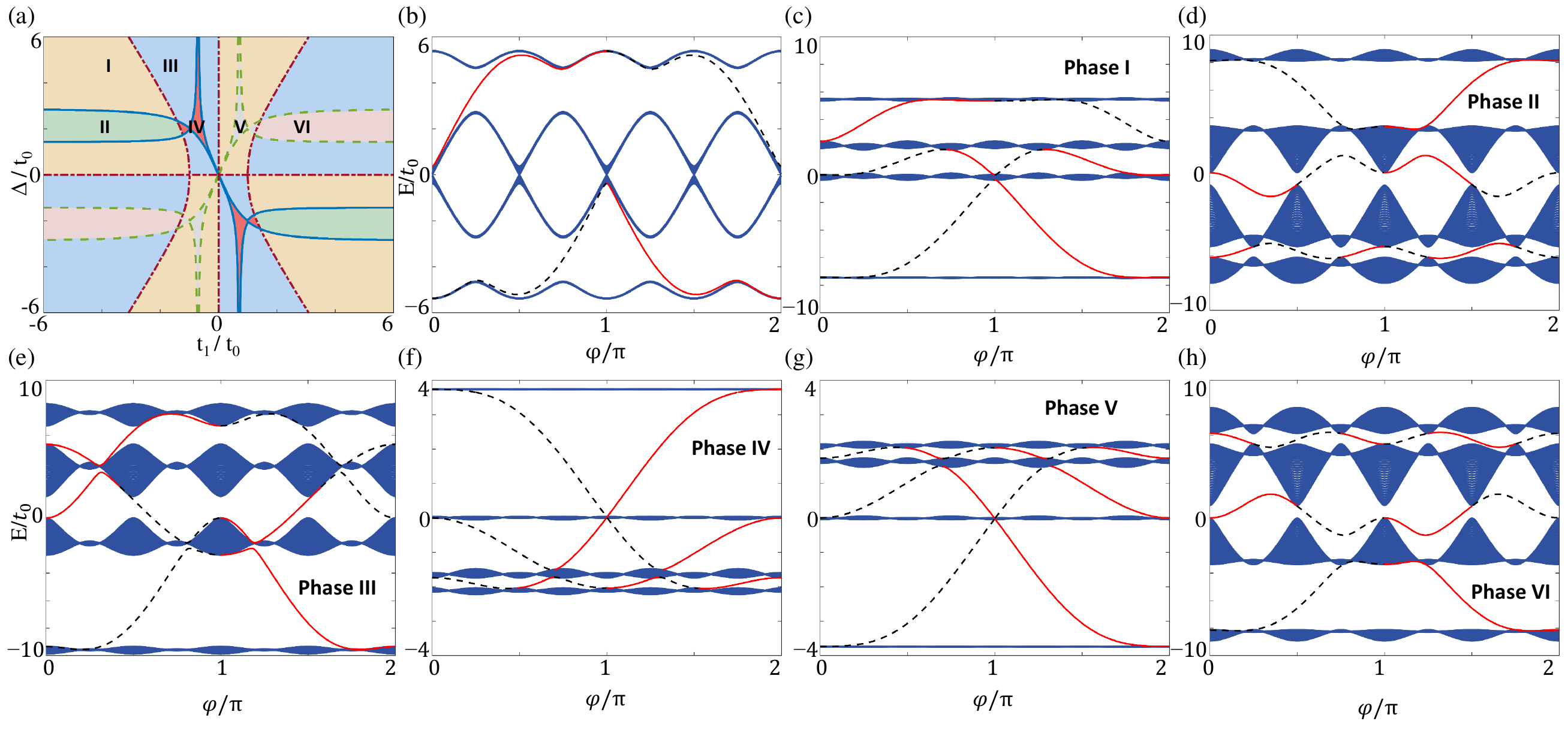}
	\caption{(a) Topological phase diagram in the parameter space of $t_1$ and $\Delta$ for $\beta=1/4$. Different topological phases are separated by the closing of the first (blue solid line), second (red dash-dotted line) or third (green dashed line) energy gaps. The energy spectra of edge states for the standard AAH model and for the generalized AAH model in the phase I-VI are shown in (b) and (c-h), respectively. The parameters are (b) $t_1=0,\Delta=5t_0$, (c) $t_1=2t_0,\Delta=5t_0$, (d) $t_1=4t_0,\Delta=-2t_0$, (e) $t_1=4t_0,\Delta=4t_0$, (f) $t_1=0.8t_0,\Delta=-2t_0$, (g) $t_1=0.8t_0,\Delta=2t_0$ and (h) $t_1=4t_0,\Delta=2t_0$. The other parameter is $t_1=t_2$. }
\label{Fig3}
\end{figure*}

\emph{Topological phase diagrams and edge states}. Now, we show that the topological properties of the generalized AAH model are quite different from the standard AAH model. The topological origin of the one-dimensional generalized AAH model also comes from a two dimensional system. We start by studying the topological properties of the energy bands in a synthetic two-dimension momentum space. Through a Fourier transformation along the genuine lattice direction, the Hamiltonian for the generalized AAH model becomes $H_{\text{GAAH}}(k_x,\varphi)$. By scanning the relative laser phase $\varphi$ from $-\pi$ to $\pi$ and employing it as a synthetic dimension, a synthetic two-dimensional momentum space is built. Similar to the standard AAH model, for $\beta=1/q$ ($q\in Z$), in the energy spectrum $E(k_x,\varphi)$ there are $q$ energy bands. The topological property for the $n$-th energy band is characterized by a synthetic Chern number, defined as
\begin{equation}
	{C_n} = \frac{1}{{2\pi}}\int_{-\pi /q}^{\pi/q} {d{k_x}} \int_{ - \pi }^\pi {d{\varphi}{F_n}\left({{k_x},{\varphi}}\right)}
\end{equation}
where ${F_n}\left({{k_x},{\varphi}}\right)$ is the Berry curvature associated with the Bloch wave function
$|\Psi_n(k_x,\varphi)\rangle$ corresponding to the $n$-th energy band. Numerically calculating the Chern numbers allows us to obtain the full topological phase diagram of $H_{\text{GAAH}}(k_x,\phi)$.

Fig. \ref{Fig2}(a) presents the corresponding topological phase diagram for an odd $q$. We take $q=3$ as an example. In this case, each unit cell has three sites, making the system support three energy bands. As shown, compared to the standard AAH model having a single kind of topological phase~\cite{Chen2012,Mei2012}, the generalized AAH model exhibits four different kinds of topological phase. According to the Chern numbers for the three energy bands (from the bottom to the top), the four topological phases are identified as $(C_1=1,C_2=-2,C_3=1)$ in the phases I, $(C_1=-2,C_2=1,C_3=1)$ in the phase II, $(C_1=-2,C_2=4,C_3=-2)$ in the phase III and $(C_1=1,C_2=1,C_3=-2)$ in the phase IV. In contrast, the standard AAH model ($t_1=t_2=0$) possess a single topological phase~\cite{Chen2012,Mei2012}, and the corresponding Chern numbers are $(C_1=1,C_2=-2,C_3=1)$, which are the same as the phase I in the generalized AAH model.

Moreover, as exhibited in Fig. \ref{Fig2}(a), the generalized AAH mode features a variety of topological phase transitions. The transitions between different nontrivial topological phases, signified by the change of Chern numbers, are accompanied by different energy gap closings. For example, the transition between the phase I and the phase II is accompanied by the closing of the first energy gap (blue solid line), as shown by the change of the Chern numbers from $(C_1=1,C_2=-2)$ to $(C_1=-2,C_2=1)$ and the invariant of the Chern number $C_3$. Since the sum of the Chern numbers for all energy bands needs to be zero, the sum of $C_1$ and $C_2$ remain unchanged when crossing the topological phase transition. Similarly, the transition between the phase II and the phase III is corresponding to the closing of the second energy gap (red dash-dotted line), as indicated by the change of the Chern numbers from $(C_2=1,C_3=1)$ to $(C_2=4,C_3=-2)$ and the invariant of their sum. While for the transition between the phase I (II) and the phase III (IV), the first and second energy gaps both close, the specific changes for the three Chern numbers are determined by the zero sum rule.

According to the bulk-edge correspondence, these nontrivial synthetic Chern numbers guarantee the appearance of topological edge states at the boundaries of the genuine dimension. In Figs. \ref{Fig2}(b), we numerically calculate the energy spectra of the one-dimension generalized AAH model in the phases I, with open boundary condition, as a function of $\varphi$. As depicted in Figs. \ref{Fig2}(c), the modes in the energy gaps are the left and right edge states, maximally localized at the left and right edges respectively. The characteristics of the edge states in which energy gap are determined by the topology of this gap. For example, for the $n$-th energy gap, its topology is characterized by the topological invariant $C^{\text{gap}}_{n}=\sum^n_{i=1}C_i$, that is related to the topology of the energy bands below this gap; Consequently, the number and group velocity of the edge states in this energy gap is respectively decided by the amplitude and sign of $C^{\text{gap}}_{n}$. As illustrated in Figs. \ref{Fig2}(b), for the first energy gap, $C^{\text{gap}}_{1}=1$, then there is one left and right edge state in this gap; While for the second energy gap, $C^{\text{gap}}_{1}=-1$, the number of the corresponding left and right edge state is same but with opposite group velocities. This bulk-edge correspondence also can be observed in the energy spectra of edge states for the phases II-IV, as shown in \ref{Fig2}(d-f) respectively.

The topological phase diagram for the generalized AAH model in the case of an even $q$ is further investigated in Fig. \ref{Fig3}(a). As displayed, the emerged topological phases and topological phase transitions turn out to be much richer. More interestingly, in this case we find a significant difference with the standard AAH model. As presented before, the middle two energy bands in the standard AAH model for an even $q$ are gapless~\cite{Chen2012,Mei2012}; While in the generalized AAH model the corresponding two bands are gapped with nontrivial topology. Specifically, for $q=4$, there are six different kinds of gapped topological phases for the four energy bands, distinguished by the Chern numbers $(C_1=1,C_2=1,C_3=-3,C_4=1)$ in the phases I, $(C_1=-3,C_2=5,C_3=-3,C_4=1)$ in the phases II, $(C_1=1,C_2=-3,C_3=1,C_4=1)$ in the phases III, $(C_1=-3,C_2=1,C_3=1,C_4=1)$ in the phases IV, $(C_1=1,C_2=1,C_3=1,C_4=-3)$ in the phases V and $(C_1=1,C_2=-3,C_3=5,C_4=-3)$ in the phases VI. As shown by the closings of the three energy gaps in Fig. \ref{Fig3}(a), this case gives rise to more topological phase transitions. The bulk-edge correspondence is studied in Fig. \ref{Fig3}(c-h). As indicated, these topological phases lead to various forms of in-gap edge states. By contrast, the standard AAH model features a single topological phase~\cite{Chen2012,Mei2012}, where the two middle energy bands are gapless and the other two gapped energy bands are topologically nontrivial with $C_1=1$ and $C_4=-1$, respectively manifested in Fig. \ref{Fig3}(b) by the central band touching and one pair of edge states in the bottom and top energy gaps.

The topological properties of the generalized AAH model can be understood by mapping it to a two-dimensional lattice model,
\begin{align}
	\hat H_{\text{GIQH}}&=\sum\limits_{m,n} [t_{0}c_{m,n}^\dag c_{m+1,n}
                +t_ye^{i2\pi\beta m}c_{m,n}^\dag c_{m,n+1}    \nonumber \\
-t^{\prime}_{y}&e^{i2\pi\beta m}(c_{m,n}^\dag c_{m+1,n+1}+c_{m,n}^\dag c_{m-1,n+1})+\text{H.c.}],
\end{align}
where $t^{\prime}_{y}=(t_1+it_2)/2$. In this mapping, we perform a Fourier transformation to the synthetic momentum $\varphi$ and transfer $H_{\text{GAAH}}(\varphi)$ into two-dimension real spaces. Compared to the lattice model in Eq. (\ref{IQH}) describing the standard IQH effect, the above lattice model supports next nearest-neighbour hoppings in the presence of magnetic fields. As a result, this model depicts a generalized IQH effect, and the corresponding topological phases in the generalized AAH model belong to the A class without respecting any symmetry.

Before summary, we briefly discuss the detection of topological phases in the generalized AAH model. As studied before, optical superlattice system constitutes an idea platform to implement quantized topological pumping~\cite{Bloch2016,Takahashi2016,Spielman2016,Bloch2018,Mei2014,Wang2013,Wei2015,Lee2016,Lee2017,Hayward2018,Mei2019,Chen2020}.
For our system, suppose it tuned into the regime where the ground band supports the topological phase with Chern number $C$, via adiabatically scanning the relative laser phase $\varphi$ over one period, one can implement a quantized topological pumping, where the displacement of the atomic cloud is equal to $C$ and detects the topological invariant. Another signatures associated with topological phases is the appearance of edge state at the boundaries of the systems. By engineering laser to create boundaries or interfaces in optical lattices and tuning the filling factor, the edge states in different energy gaps and their dynamics can be directly observed through time-of-flight~\cite{DE1,DE2} or Bragg spectroscopy technology~\cite{DE3}.

\emph{Summary and Outlook}. In summary, we have shown that the optical superlattice system in the limiting case $V_1\gg V_2$ implements the generalized AAH model, with periodic modulations on both the on-site energies and nearest-neighbour hoppings. By calculating topological invariants and edge states, we have exhibited that this system supports much richer topological properties, which are absent in the standard AAH model.

As shown, the generalized AAH model provides opportunities to go beyond what is possible in the standard AAH model. For example, in the near future it would be quite interesting to generalize our result to two dimensions for implementing two-dimensional generalized AAH model. Through introducing the relative laser phases in the two directions as two synthetic momentums, this model not only enables the implementation of four-dimensional topological phases~\cite{Zilberberg2013,Goldman2015}, but also allows us to explore a variety of four-dimensional topological phases transitions that are quite challenging before. Moreover, the generalized AAH model in the incommensurate case also can set an stage for controlling Anderson localization~\cite{AA,Ganeshan2015}.

\emph{Acknowledgment}. This work was supported by the National Key Research and Development Program of China (2017YFA0304203), National Natural Science Foundation of China (NSFC) (12034012, 12074234), Changjiang Scholars and Innovative Research Team in University of Ministry of Education of China (PCSIRT)(IRT\_17R70), Fund for Shanxi 1331 Project Key Subjects Construction, and 111 Project (D18001).

\end{document}